\documentstyle[prd,aps,floats,graphicx]{revtex}

\begin{document}
\draft

\twocolumn[\hsize\textwidth\columnwidth\hsize\csname
@twocolumnfalse\endcsname

\title{\bf Quintessence models in Supergravity}

\author{E.~J.~Copeland,\cite{ejc}~ N.~J.~Nunes,\cite{njn}~ and F.~Rosati\cite{fr}}

\address{Centre for Theoretical Physics, University of Sussex,
Falmer, Brighton BN1 9QJ, United Kingdom }

\date{\today}

\maketitle

\begin{abstract} 
Scalar field models of quintessence typically require that
the expectation value of the field today is of order the Planck mass, 
if we want them to explain the observed
acceleration of the Universe.
This suggests that we should be considering models in the context 
of supergravity. We discuss a particular class of supergravity 
models and analyze their behavior
under different choices of the K\"ahler metric.
\end{abstract}
\pacs{PACS numbers: $98.80.$Cq \hspace*{1.5cm} hep-ph/0005222 \hspace*{1.5cm} 
 SUSX-TH/00-008}
\vskip2pc]

\section{Introduction}

Several cosmological observations appear to suggest that the present Universe is
dominated by an unknown form of energy with negative pressure, acting as an
effective cosmological constant $\Lambda$.
CMB and cluster distribution data, when combined with the measurements of the
redshift--luminosity relation using high redshift type Ia supernovae, single out
a flat cosmological model with $\Omega_{\rm matter} \sim 1/3$ and $\Omega_{\Lambda} 
\sim 2/3$ \cite{data}.
The contribution to the energy density which has led to the observed
acceleration is named ``quintessence'' and  could well be some more general
component than a bare cosmological term $\Lambda$ \cite{quint}.
In particular, a scalar field $Q$ rolling down a potential $V(Q)$ can act as an
effective dynamical cosmological constant with equation of state
\begin{equation}
w_{Q} \ = \ \frac{p_{Q}}{\rho_{Q}} \ = \
\frac{\dot{Q}^2/2 - V(Q)}{\dot{Q}^2/2 + V(Q)} \;\; .
\end{equation}

Dynamical models are appealing because they could potentially solve the smallness
and coincidence problems which arise when postulating that there is today
a cosmological constant energy density
contribution, $\rho_{\Lambda} = \Lambda /8\pi G$, of the same order of magnitude
as the critical energy density $\rho_c \simeq 10^{-47} {\rm GeV^4}$.
The advantage of dealing with a dynamical scalar field is
that there exist attractor solutions in a cosmological setting for some classes
of potentials \cite{scalcosmo}.
If this is the case, the scalar field will join the attractor solution before
the present epoch for a very wide range of initial conditions, thus relieving
the fine--tuning issue.

For example, the potential $V(Q) = M^{4+\alpha} Q^{-\alpha}$,
with $\alpha > 0$, leads to a `tracking' attractor solution for which
the scalar energy density scales as $\rho_{Q}/ \rho_B \sim a^{3(w_B - w_{Q})}$,
where $\rho_B$ and $w_B$ are the background energy density and equation of state
(the latter being $=0, 1/3$ during matter and radiation domination respectively),
and $a$ is the scale factor of the Universe.
The equation of state of the attractor is given by
$w_{Q} = (w_B \alpha -2)/(\alpha +2)$  and is always negative during
matter domination \cite{track}.
If $Q$ has already reached the attractor solution today, it  must be of order
of the Planck mass $M_{\rm Pl}$.
This follows from the fact that for the tracking solutions $V^{\prime\prime}(Q)
\simeq H^2$ and moreover we require $V \simeq \rho_c$ today.
The scale $M$ in the potential is fixed by
requiring that $V(Q \simeq M_{\rm Pl}) \simeq \rho_c$ and depends on the exponent 
$\alpha$.
This class of potentials is then a suitable candidate for quintessence and has
been shown to be compatible with observations \cite{track}.

If instead an exponential potential is considered, $V(Q) = M^4\,
\exp{(\lambda Q)}$, two attractors are found \cite{scale}, 
depending on the value of $\lambda$.
If $\lambda > 3(w_B+1)$, the late time attractor is a `scaling' one, characterized
by $w_{Q} = w_B$ and $\Omega_{Q} = 3(w_B +1)/ \lambda^2$.
Note that in this case the ratio of the scalar to critical energy density is
independent of the mass scale $M$ in the potential. Unfortunately the equation
of state is non negative and so cannot drive the Universe to an accelerated 
expansion.
If $\lambda < 3(w_B +1)$, instead, then the late time attractor is the scalar field 
dominated
solution with $\Omega_{Q} =1$ and $w_{Q} = -1+ \lambda^2 /3$.
Now the equation of state is negative, but clearly cannot be the solution for all 
times,
as there is the tight constraint on the allowed contribution of the scalar field at
nucleosynthesis of $\Omega_Q(1 \rm{MeV}) < 0.13$. 
Besides, one must allow time for structure formation 
before the Universe starts accelerating. A scenario like this requires that the 
scalar
field starts with roughly the same energy density as today, and so provides no 
improvement
on solving the fine tuning problem.
The exponential potentials, then, cannot be used individually for modeling
quintessence, but interesting modifications have been proposed in order to fit the 
data
\cite{nelson,otherexp}.

\subsection{Particle physics models}

While dealing with what is thought to be one of the fundamental components of the
present universe, it is mandatory to ask if we can find any firmer motivation
for introducing a cosmological scalar field.
We actually know that scalar fields arise quite naturally in unified theories and
are a necessity in supersymmetric theories where they play the role of 
`superpartners'
of the Standard model fermion fields.
Scalar fields are also required by another crucial cosmological mechanism,
inflation. Indeed, it has been recently shown \cite{qinfl} that the same unique field 
could
possibly dominate both the inflationary and quintessence phases of our universe.
The issue of finding deeper roots for the quintessence scalar into `realistic'
particle physics models is then a pressing one.

It is well known that inverse power law scalar potentials arise in 
supersymmetry (SUSY) gauge
theories due to non--perturbative effects \cite{SQCD}.
They have been studied as candidates for quintessence \cite{smodel} and shown to
provide a viable phenomenology both in the one--scalar and multi--scalar cases.
Exponential potentials are also common in high energy physics, since they arise as a
consequence of Kaluza--Klein type compactifications of string theory.
The attempt of building a particle physics motivated model for quintessence
is then a well-posed problem, although many points remain to be clarified.
For example, exponential-like models as those proposed in \cite{nelson,otherexp}
still await a firmer high energy motivation.
At the same time, inverse power law models based on SUSY QCD need to be
revisited taking into account the possibility that the scalar potential
receive corrections originating from quantum or supergravity effects.

This last issue was recently addressed by Brax and Martin \cite{brax}.
In particular they showed that inverse power law models which lead to quintessence
are stable against quantum corrections, both in the supersymmetric and
non--supersymmetric cases.
Furthermore, discussing curvature and K\"ahlerian corrections in the globally
supersymmetric case, they found that while curvature corrections are
negligible, K\"ahlerian ones  might be important in the early evolution of $Q$.

The most interesting question, though, concerns supergravity (SUGRA)
corrections \cite{brax}.
Since today the vacuum expectation value (vev) of the scalar field is of order 
Planck mass, these corrections are
unavoidable and may lead to problems for these power law quintessence models.
In particular, if the scalar potential is dramatically altered 
for field values corresponding
to the most recent epoch, then we might have a completely different 
phenomenology
from the attractor case just described.
In ref. \cite{brax}, a possible way of avoiding the dangerous SUGRA corrections
was outlined, which involved the restrictive condition $\langle W \rangle = 0$.
The aim of this paper is to further develop this line of attack by proposing an
alternative way of tackling SUGRA corrections.
As an example, we will study the cosmology of a specific SUGRA--inspired
quintessence model and show that it is compatible with observations for 
a wide range of initial conditions.

\section{Supergravity models}

\subsection{Global supersymmetry}

Globally supersymmetric QCD theories with $N_{c}$ colors and $N_{f}<N_{c}$
flavors  may give an explicit realization of a model of quintessence with an
inverse power law scalar potential \cite{smodel}.
The matter content of the theory is given by the chiral superfields
$Q_{i}$ and $\overline{Q}_{i}$ ($i=1\ldots N_{f}$) transforming
according to the $ N_{c}$ and $\overline{N}_{c}$ representations of
$SU(N_c)$, respectively.

Supersymmetry and anomaly-free global symmetries constrain the
superpotential to take the form
\begin{equation}
W=(N_{c}-N_{f})\left( \frac{\Lambda ^{(3N_{c}-N_{f})}}{{\rm
det}T}\right) ^{ \frac{1}{N_{c}-N_{f}}} \label{superpot}
\end{equation}
where the gauge-invariant matrix superfield $T_{ij}=Q_{i}\cdot
\overline{Q}_{j}$ appears. $\Lambda $ is the only mass scale of the
theory.
It is the supersymmetric analogue of $\Lambda _{QCD}$, the
renormalization group invariant scale at which the gauge coupling of
$SU(N_{c})$ becomes non-perturbative.

If the K\"ahler metric is flat, the scalar potential is given by\footnote{In
the following, the same symbols will be used for the superfields $Q_{i}$,
$\overline{Q}_{i}$, and their scalar components.}
\begin{eqnarray}
V(Q_{i},\overline{Q}_{i}) &=& V_F + V_D \nonumber \\ &=&
\sum_{i=1}^{N_{f}}\left( |F_i|^{2}+|F_{\overline{i}}|^{2}\right)
+\frac{1}{2}D^{a}D^{a} \label{potscal}
\end{eqnarray}
where $F_i=\partial W/\partial Q_{i}$,
$F_{\overline{i}}=\partial W/\partial \overline{Q}_{i}$, and
\begin{equation}
D^{a}=Q_{i}^{\dagger }t^{a}Q_{i}-\overline{Q}_{i}t^{a}\overline{Q}
_{i}^{\dagger }\;\; ,  \label{d-terms}
\end{equation}
with the $t^a$'s being the generators of the gauge group.
Of interest to us are the  dynamics of the field expectation values
which take place along directions in field space in which the 
above D-term vanishes,
{\it i.e.} the perturbatively flat directions $\langle Q_{i\alpha }\rangle
=\langle \overline{Q}_{i\alpha }^{\dagger }\rangle $, where $\alpha
=1\cdots N_{c}$ is the gauge index.
At the non-perturbative level these directions acquire a non vanishing
potential from the F-terms in (\ref{potscal}), which are zero to any order
in perturbation theory.
Gauge and flavor rotations can be used to diagonalize the
$\langle Q_{i\alpha }\rangle $ and put them in the form
\[
\langle Q_{i\alpha }\rangle =\langle \overline{Q}_{i\alpha }^{\dagger}\rangle =
\left\{
\begin{array}{ll}
Q_{i}\delta _{i\alpha } \hspace{5mm} 1 \leq \alpha \leq N_{f} \\
0 \hspace{9mm} N_{f} \leq \alpha \leq N_{c}
\end{array}
\right. \,.
\]
Along these directions, if the expectation values of all the $N_f$ scalars
are taken to be equal, $\langle Q_i \rangle = Q, \; i= 1,\dots , N_f$,
the cosmological evolution of the scalar vev $Q$ is given by
\begin{eqnarray}
\ddot{Q} &=& -3H\dot{Q} + \beta
\frac{\Lambda^{4+2\beta}}{Q^{2\beta +1}} ~ , \nonumber \\
\beta &=& \frac{N_{c}+N_{f}}{N_{c}-N_{f}}\ ,
\end{eqnarray}
thus reproducing exactly the case of a single scalar field $Q$ in a
potential $V=\frac{\Lambda ^{4+2\beta}}{2}\, Q ^{-2\beta}$.
The `tracker' solution is characterized \cite{smodel} by an equation of state
\begin{equation}
w_Q \ =\ \frac{N_c+N_f}{2N_c}\ w_B \ -\ \frac{N_c-N_f}{2N_c}
\end{equation}
as a function of the parameters of the theory, with the scalar energy density
growing with respect to the matter one as
\begin{equation}
\rho_Q / \rho_{\rm m} \ = \ a^{3(N_c-N_f)/2N_c} \;\; .
\end{equation}
Requiring that the scalar $Q$ both has reached the tracker today and is starting
to dominate the energy density, we obtain that at the present epoch
$Q \simeq M_{\rm Pl}$ and that the mass scale in the potential must satisfy
$\Lambda^{4+\beta} \simeq \rho_c ~ M_{\rm Pl}^{\beta}$, introducing a 
degree of fine tuning
in the problem.

\subsection{Supergravity corrections}

The above discussion is valid as far as the global SUSY limit can be taken
as a good approximation. This is correct for most of the cosmological evolution
of the scalar $Q$.
However, for the present epoch we are not allowed to neglect SUGRA corrections,
since we enter the regime for which they might become important.
In this case, the F-term in the scalar potential in general is
\begin{eqnarray}
 V(Q) &=& F^2 - e^{\kappa^2K} 3\kappa^2|W|^2  \nonumber  \\
 &=& e^{\kappa^2K} [( W_i +\kappa^2 W K_i ) K^{j^*i} 
( W_j + \kappa^2 W K_j )^* \nonumber \\
&-& 3\kappa^2|W|^2] 
\label{fterm}
\end{eqnarray}
where the subscript $i$ indicates the derivative with respect to
the $i$-th field, and $\kappa^2 = 8\pi G=8\pi M_{\rm Pl}^{-2}$.

Brax and Martin \cite{brax} discuss the  case of a theory with
superpotential $W=\Lambda^{3+\alpha} Q^{-\alpha}$ and a flat
K\"ahler potential, $K = QQ^*$.
It is straightforward to compute the resulting scalar potential:
\begin{equation}
V(Q) = e^{\frac{\kappa^2}{2}Q^2}  \frac{\Lambda^{4+\beta}}{Q^{\beta}} \,
\left( \frac{(\beta -2)^2}{4} - (\beta +1) \frac{\kappa^2}{2}Q^2 + 
\frac{\kappa^4}{4}Q^4 \right) 
\label{flatpot}
\end{equation}
where $\beta = 2\alpha +2\ $.
The main effect of the supergravity corrections is that the scalar potential can
now become negative due to the presence of the second term.
This is a serious drawback for the model, which becomes ill defined for
$Q \simeq M_{\rm Pl}$.
They go on to propose a possible solution by imposing the condition
that the expectation value of the superpotential
vanishes, $\langle W \rangle =0$. We then see from equation (\ref{fterm}) that the
negative contribution to the scalar potential disappears, and it takes the form 
\begin{equation}
V(Q) \ =\ \frac{\Lambda^{4+\alpha}}{Q^{\alpha}}\,  e^{\frac{\kappa^2}{2}Q^2} \;\; .
\end{equation}
The condition $\langle W \rangle =0$ is possible to realize, 
for example, in a model in which we allow 
matter fields
to be present in addition to the quintessence scalar field \cite{brax}.
Then, if at least one of the gradients of the superpotential with respect 
to the matter fields
is non--zero, the scalar potential will always be positive.

This is not the only possibility, though.
There are two obvious problems with the potential (\ref{flatpot}): 
one, as already stated, is the negative term in the general 
expression (\ref{fterm}) and the second is the choice of the
K\"ahler metric which makes this term grow with the field's vev, relative
to the other terms in the potential.
As mentioned above, setting $\langle W \rangle =0$ is a tight restriction, so we
will address the issue by relaxing this constraint but allow for more 
general forms of the K\"ahler metric. 

Such an approach was recently 
adopted in \cite{casas96,binetruy97} as a method of obtaining a minimum 
for the dilaton field in string theory.
It had the advantage of relying on only one gaugino condensate and
provided an alternative approach to the phenomenology associated 
with `racetrack' models \cite{decarlos93}. 
In this
scenario, the K\"ahler potential acquires string inspired non-perturbative 
corrections.
A further nice feature of these models is that it is possible to 
have a minimum with zero
or small positive cosmological constant \cite{binetruy97,barreiro98}, 
and moreover it is possible to
establish that the dilaton can be stabilized in such a minimum 
in a cosmological setting
\cite{barreiro98a}. 

In general, for different choices of the K\"ahler metric, the negative 
term in (\ref{fterm})
does not always lead to the disaster of a negative minimum in 
the scalar potential.
For a general K\"ahler, we do not know {\it a priori}  the 
shape of the potential or the location of the minimum.
In fact, in what follows we will show through 
explicit examples that the scalar
potential might always remain positive through a 
suitable choice of the K\"ahler metric.
Moreover, with this approach there is no need to introduce additional fields
in the model.

\vskip0.15in

Let us now go on to study  SUGRA corrections to inverse power law 
quintessence models by choosing more general K\"ahler potentials.
Consider, for example, a theory with superpotential
$W = \Lambda^{3+\alpha}\, \tilde{Q}^{-\alpha}$
and a K\"ahler $K = -\ln (\kappa \tilde{Q} + \kappa \tilde{Q}^*)/\kappa^2 \, $, 
the type of term which is present at tree level for both
the dilaton and moduli fields in string theory.
In this case, the resulting scalar
potential, expressed in terms of the canonically normalized field 
$Q = (\ln \kappa\tilde{Q})/\sqrt{2} \kappa$, is
\begin{equation}
V(Q) = M^4~e^{-\sqrt{2}\beta \kappa~ Q}
\end{equation}
where $M^4 = \Lambda^{5+\beta} ~\kappa^{1+\beta}~(\beta^2 -3)/2 $ with
$\beta = 2\alpha +1  >  \sqrt{3}$ to allow for positivity of the potential.
This corresponds to the `scaling' solution discussed in the introduction and so
cannot lead to a negative equation of state for the field in a matter 
dominated regime.

Another example follows as a natural extension of the one just described 
and leads to potentials with more than one exponential. 
For a superpotential of the form
$W = \Lambda^{3+\alpha}\, \tilde{Q}^{-\alpha} + 
\Lambda^{3+\beta}\, \tilde{Q}^{-\beta}$
and the same K\"ahler metric as above, then in terms of the same 
canonically normalized
field $Q = (\ln \kappa\tilde{Q})/\sqrt{2}\kappa$, the scalar potential becomes
  
\begin{eqnarray}
V(Q) &=& (M_1)^4~ e^{-\sqrt{2}\gamma \kappa\, Q} + 
(M_2)^4~ e^{-\sqrt{2}\delta \kappa\, 
Q}
\\ \nonumber
&+& (M_3)^4~ e^{- \frac{\gamma + \delta}{\sqrt{2}} \kappa\, Q} \,,
\end{eqnarray}
where $\gamma = 2\alpha + 1$, $\delta = 2\beta+1$ and 
\begin{eqnarray}
(M_1)^4 &=& \Lambda^{5+\gamma} ~\kappa^{1+\gamma}~(\gamma^2 -3)/2  \,, \\ \nonumber
(M_2)^4 &=& \Lambda^{5+\delta} ~\kappa^{1+\delta}(\delta^2 -3)/2  \,, \\ \nonumber
(M_3)^4 &=& \Lambda^{5+\frac{\gamma+\delta}{2}} ~\kappa^{1+\frac{\gamma + \delta}{2}}
          ~(\gamma \delta -3)  \,.
\end{eqnarray}

At first sight this appears to be of the form required in \cite{nelson} 
in that it involves
multiple exponential terms. However, closer analysis indicates that 
the slopes of the exponentials are not adequate to satisfy the 
bounds arising from nucleosynthesis constraints, whilst also providing 
a positive cosmological constant type contribution today.

As we mentioned earlier, it is possible to have more general
K\"ahler potentials, and with that in mind we now consider the
original model $W = \Lambda^{3+\alpha}\, \tilde{Q}^{-\alpha}$, but
with a K\"ahler potential which depends on a parameter $\gamma$
\begin{equation}
K = \frac{1}{\kappa^2}~\left[ \ln (\kappa\tilde{Q} + 
\kappa\tilde{Q}^*) \right]^\gamma \,,
\hspace{1cm}  \gamma > 1 \;\; .
\end{equation}
In this case, the second derivative of the K\"ahler is
\begin{eqnarray}
K_{\tilde{Q}\tilde{Q}^*} &=& \frac{\gamma (\gamma -1)}{\kappa^2 
~(\tilde{Q}+ \tilde{Q}^*)^2} \
[ \ln (\kappa \tilde{Q}+ \kappa \tilde{Q}^*)]^{\gamma -2} \, \\ \nonumber  
&\times& \left( 1- \frac{\ln (\kappa\tilde{Q}+ \kappa \tilde{Q}^*)}{\gamma -1} 
\right)
\label{kinetic}
\end{eqnarray}
and the canonically normalized field $Q$ can be obtained as a 
function of $\tilde{Q}$
by integrating the following expression
\begin{equation}
d Q \ =\ \sqrt{2 \, K_{\tilde{Q}\tilde{Q}^*}}\ d\tilde{Q} \;\; .
\end{equation}
In order to avoid the singularity at 
$\tilde{Q}+\tilde{Q}^*=1/\kappa^2$, 
when $\ln (\kappa\tilde{Q} + \kappa\tilde{Q}^*)$
passes through zero (see equation (\ref{kinetic})), the only possible choice 
is $\gamma =2$.
We then obtain:
\begin{equation}
K_{\tilde{Q}\tilde{Q}^*} \ =\ \frac{2\, 
[1- \ln (\kappa\tilde{Q}+\kappa\tilde{Q}^*)]}
{\kappa^2 (\tilde{Q}+\tilde{Q}^*)^2}
\end{equation}
and as a consequence
\begin{equation}
Q \ =\
-\frac{2}{3\kappa} \, [1-\ln (2\kappa\tilde{Q})]^{3/2} \;\; .
\end{equation}
Implying  that the theory is well defined for 
\[
-\infty < \ln (2\kappa\tilde{Q}) < 1
\]
which corresponds to $0< \tilde{Q} <e/2\kappa \, $. 

The scalar potential in terms of the canonically normalized field $Q$ reads
\begin{eqnarray}
V &=& M^4 ~ \left[ 2x^2 +(4\alpha -7)x+2(\alpha -1)^2
\right]  ~ \frac{1}{x} \nonumber \\
&~&  \times  \exp [(1-x)^2 -2\alpha (1-x)]
\label{vpot}
\end{eqnarray}
where for notational convenience we have defined the quantities
\begin{equation}
x \  \equiv  \left( -\frac{3}{2}\, \kappa Q \right)^{2/3} \ =\ 
1-\ln (2 \kappa \tilde{Q})  \,, 
\label{x-q}
\end{equation}
\begin{equation}
M^4 = \Lambda^{6+2\alpha}~ \kappa^{2+2\alpha}~ 2^{2\alpha} \, . 
\end{equation}
Note that the canonically normalized field $Q$ has a range $-\infty <Q<0$.

We can see from equations (\ref{vpot})--(\ref{x-q}) 
that the scalar potential behaves like
an exponential for $|Q| \gg 1$ and like an inverse power law for
$|Q| \ll 1$, and thus develops a minimum at a finite value $Q_{\rm m}$.
Note that the potential is always positive for any $\alpha > 1.25$.
Thus, we have found that in this case the supergravity corrections induce a 
finite minimum in the potential but do not spoil its positivity.
Note also that the field's value in the minimum is exactly 
in the region where we
expect the supergravity corrections to become relevant. 
For example, with $\alpha =5$
we obtain $Q_{\rm m} \simeq -0.02$ (in $8 \pi G =1$ units),  which corresponds to
$\tilde{Q} \simeq 1.2$.
Imposing that the minimum of the potential equals the critical energy density
today, we can also estimate the mass scale $\Lambda$, depending on $\alpha$.
In the case $\alpha = 5$ we have that $V(Q = Q_{\rm m}) 
\simeq 10^{-47}\rm{GeV}^4$
which corresponds to $\Lambda \simeq 6~10^{10}\, {\rm GeV}$.

\subsection{Supersymmetry breaking}

If supersymmetry is to be realized in nature, it must be broken at a 
mass scale $M_S$ such that $M_S^2 \sim \langle F \rangle \gtrsim (10^{10} 
{\rm GeV})^2$ or   $M_S^2 \sim \langle F \rangle \gtrsim (10^{4} 
{\rm GeV})^2$ (for gravity and gauge mediated cases respectively), 
in order to lift the supersymmetric scalar particle masses
 above $10^2 {\rm GeV}$.

  This then requires the superpotential in (\ref{fterm}) to be 
$W \sim\langle F \rangle \kappa^{-1} \sim m_{3/2} \kappa^{-2}$ in order to cancel
the F-term contribution and consequently to give a negligible total vacuum 
energy density ($m_{3/2}$ is the gravitino mass).

  From the discussion in the last section\footnote{We thank 
K.~Choi and D. Lyth for discussions on this point.}, 
it is clear that the dynamical
cosmological constant provided by the quintessence potential cannot
be the dominant source of SUSY breaking in the Universe as 
$W \simeq \Lambda^{3+\alpha} \kappa^{-\alpha}\sim (10^{-3} {\rm eV})^2 \kappa^{-1} \ll 
\langle F \rangle \kappa^{-1}$. Therefore,
we need some additional source of SUSY breaking. If we consider now 
the superpotential,
\begin{equation}
W =  \Lambda^{3+\alpha}~Q^{-\alpha} + m_{3/2} \kappa^{-2} \,,
\end{equation}
then one gains a correction to the scalar potential in (\ref{vpot}) of,
\begin{equation}
\delta V \sim \Lambda^{3+\alpha} m_{3/2} \kappa^{-\alpha} + 
m_{3/2}^2 \kappa^{-2} \,,
\end{equation}
for  $Q \sim \rm{M_{Pl}}$ today \cite{choi}.

  The first term can in principle be controlled for sufficiently
 large $\alpha$, however,
the constant second term unavoidably leads to a disruption of the quintessence 
potential. This is a very serious problem of all supergravity models 
in quintessence \footnote{In order to avoid the SUGRA corrections 
problem, Choi proposes a Goldstone-type quintessence model 
inspired in heterotic M-Theory \cite{choi}.}.

The situation gets worse, since, as pointed out in \cite{lyth}, 
it appears that even if we 
imagine that the amount of SUSY breaking that we observe in the universe today 
comes from a hidden sector other than the quintessence one, there will 
still be gravitational couplings between the two sectors that rekindle 
the original problem. 

However, some recent proposals point in a slightly different 
direction for solving 
this problem. The basic idea is that the traditional approach 
to SUSY breaking might 
not be the best way to explain the world we live in. 
For example, the mass difference between the superpartners could 
arise in a 4D world 
with {\it unbroken} SUSY through some higher dimensional effects 
\cite{witten}. If 
this is the case, we would not need to break supersymmetry, and 
the quintessence 
potential would be preserved. 
Another possibility \cite{banks} is that the relation between 
the SUSY breaking scale 
$M_S$ and the cosmological constant $\langle F \rangle ^2$ is not what is 
usually considered.
If $M_S =  \kappa^{-1} [ \langle F \rangle ^2 {\kappa}^{4}]^{\beta}$ 
and $\beta = 1/8$, 
instead of the 
usual $1/4$, then the observed cosmological constant would provide 
just the right 
amount of SUSY breaking. In this case we wouldn't have any dangerous 
F-term of order 
$\sim \kappa^{-2} M_S^2$ which spoils the quintessence potential.

\section{Cosmology of the model}
We can now study in further detail the model we presented in the previous 
section.
It can be easily checked that for $Q \rightarrow 0$ the potential 
(\ref{vpot}) goes as
$V \sim Q^{-2/3}$, while for $Q \rightarrow -\infty$ it is $V \sim
Q^{2/3} e^{Q^{4/3}}$.
This behavior is {\em independent} of the parameter $\alpha$ in our theory, 
which
plays no role in the asymptotic form of the potential.
For all the values of $\alpha$ we then find the same qualitative behavior 
for the scalar
$Q$. For a very wide range of the initial conditions, indeed, 
we obtain scalar field dominance and negative equation of 
state at the present epoch.

\begin{figure}[hb!]
\includegraphics[height=6cm,width=8cm]{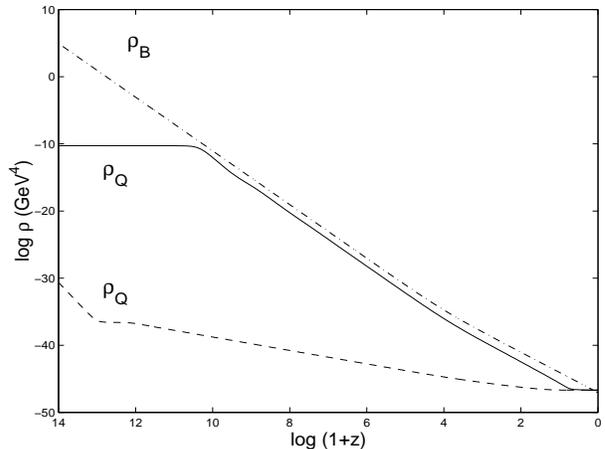}
\caption{  The evolution of
the energy densities $\rho$ of different cosmological components is given
as a function of redshift, for the case $\alpha =5$ and with $\Omega_Q({\rm today})
\simeq 0.7$.
The dot-dashed line represents the background energy density
$\rho_B = \rho_{\rm radiation} + \rho_{\rm matter}$.
The solid line is the evolution of the scalar field energy density when the field $Q$
starts from an initial value $Q_{\rm in}/Q_{\rm m} \gg 1$, while the dashed line 
corresponds to starting with $Q_{\rm in}/ Q_{\rm m} \ll 1$.}
\end{figure}

In Figure 1 and Figure 2 we plot the evolution of the scalar
energy density and equation of state for $\alpha = 5$. 

If the scalar field is rolling down the potential towards the minimum
from the  side $Q_{\rm in}/Q_{\rm m} \ll 1$, then it will exhibit
a `tracking' behavior as in the general case $V \sim Q^{-\beta}$, with $\beta =2/3$.
This is characterized by an equation of state
$w_Q = (w_B \beta -2)/(\beta +2) = -2/3, -3/4$ during radiation and matter domination
respectively.
When the field $Q$ approaches the minimum, it will depart from the attractor solution
and enter a regime of damped oscillations.
The non--zero vacuum energy will rapidly take over the kinetic term and the equation
of state be driven towards $w_Q =-1$ (see Figure 2).

If, instead, $Q$ rolls down from the side $Q_{\rm in}/Q_{\rm m} \gg 1$ the 
exponential in (\ref{vpot}) is then important. 
As it will be shown below, this leads to an attractor with $w_Q \simeq w_B$ 
up to a scale factor dependent logarithmic correction. 
Eventually, the field will settle down in the minimum with $w_Q = -1$,
after a stage of small oscillations about the minimum as before.
The earlier requirement made on $\alpha$, namely $\alpha > 1.25$ is also sufficient 
to respect the nucleosynthesis bound of $\Omega_Q(1 \rm{MeV}) < 0.13$.

\begin{figure}[ht]
\includegraphics[height=6cm,width=8cm]{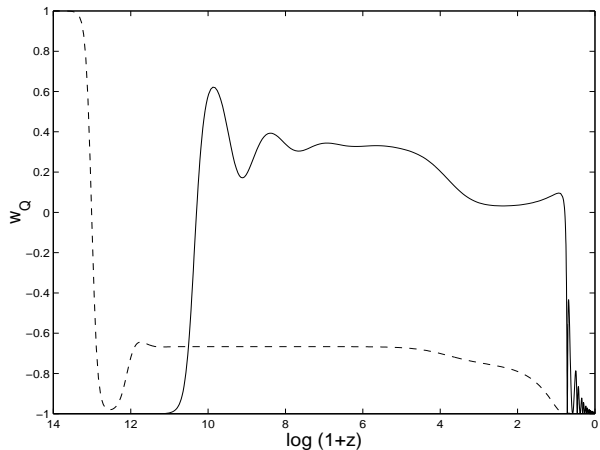}
\caption{ The evolution of the scalar field equation of state
as a function of redshift, corresponding to the two cases of FIG 1.}
\end{figure}

In the second reference of \cite{track} a general formula for the scalar 
equation of state in the case of `tracking' potentials was derived:
\begin{equation}
w_Q ~ \simeq ~ \frac{w_B -2~ (\Gamma -1)}{1+2~(\Gamma -1)} \;\; ,
\label{zws}
\end{equation}
where $\Gamma$ is defined as $\Gamma \equiv V^{\prime\prime}V/(V^{\prime})^2$,
with a prime denoting the derivative with respect to the scalar $Q$.

In deriving eq.~(\ref{zws}), it is assumed that $\Gamma$ is nearly constant. 
A tracker solution with $w_Q < w_B$ is reached if $\Gamma > 1$.
For a potential of the form 
\begin{equation}
V = M^4 e ^{\alpha (\kappa Q)^\beta} \;\; , 
\end{equation}
we obtain
\begin{equation}
\Gamma = 1 + \frac{\beta -1}{\alpha \beta} ~ \frac{1}{(\kappa Q)^\beta} \;\; ,
\label{gamma}
\end{equation}
which is larger than 1 and slowly varying as required. Actually the dependence of 
$\Gamma$ on the scale factor $a(t)$ through $Q(a)$ will give the logarithmic 
variation of the equation of state mentioned earlier. 

By substituting the expression (\ref{gamma}) for $\Gamma$ in eq.~(\ref{zws}), 
we obtain
\begin{equation}
w_B = w_Q + (w_Q +1) ~ \frac{2(\beta -1)}{\beta} ~ \frac{1}{(\kappa Q)^\beta} \;\; ,
\label{w}
\end{equation}
with 
\begin{equation}
(\kappa Q)^\beta = \ln \left( \frac{V_0}{M^4}\right)  - 3(w_Q +1)~ 
\ln \left( \frac{a}{a_0} \right) \;\; ,
\end{equation}
where $V_0$ is the value of the scalar field potential at some time $a=a_0$,
when the attractor has already been reached, and we have assumed the following 
dependence of the potential on the scale factor
\begin{equation}
V(a) = V_0 \left( \frac{a}{a_0} \right)^{-3(w_Q +1)} 
\end{equation} 
which is exact if $w_Q$ is constant and can be taken as a good approximation 
when $w_Q$ is slowly varying. 

Note that, when $\beta =1$, then $V$ reduces to a simple exponential and 
from (\ref{w}) we recover the  `scaling' behaviour with  $w_Q = w_B$.

\section{Conclusions}
In this paper we have studied supergravity corrections to
quintessence models with a superpotential $W = \Lambda^{\alpha+3}
Q^{-\alpha}$. 
The motivation for this is simple. 
Scalar field models of quintessence typically require that
the expectation value of the field today is of order the Planck mass, 
if we want it to explain the observed acceleration of the Universe.
This suggests that we should be considering models in the context 
of supergravity.

We have proposed a new line of attack to a serious problem that these
models share, {\it i.e.} the negativity of the minimum in the resulting scalar 
potential. Allowing for nonlinear modifications to the K\"ahler 
potential this problem can effectively be cured. 
Such modifications are, as far as we are aware, perfectly acceptable and 
lead to some interesting features. 

In particular we  discussed in detail a model with K\"ahler 
potential $K=[\ln (\kappa Q+ \kappa Q^*)]^2/\kappa^2$ and superpotential
 $W = \Lambda^{\alpha+3} ~Q^{-\alpha}$ . 
In this case, the minimum of the resulting scalar potential is always 
positive for any $\alpha > 1.25 $ and is located close to the point where
the SUGRA corrections start to be important, {\it i.e.} at $Q \sim M_{\rm Pl}$.
We found that the resulting scalar potential yields two possible attractor 
solutions, both of which lead the scalar field towards the minimum 
without dominating the Universe dynamics before today.
After reaching the minimum, the scalar field will mimic a cosmological constant with
$w_Q \simeq -1$.

We also noted that the dynamical cosmological constant described above cannot
be the dominant source of SUSY breaking in the universe, since the 
F-term is not large enough.
This means that we need some additional source of SUSY breaking, which 
in turn implies we must ensure that the SUSY breaking sector does not interact
with the quintessence sector, in order not to spoil the behaviour described above.
Unfortunately, this particular problem still remains to be resolved 
\cite{lyth}, without invoking some extra symmetry principle 
in the action \cite{choi}.  A possible, but still speculative resolution
has been proposed by Witten \cite{witten}, who has suggested that we may 
actually live in a vacuum with unbroken supersymmetry.  As an alternative,
Banks \cite{banks} has proposed amechanism through which the measured cosmological constant 
might well be able to provide the observed amount of supersymmetry breaking.

In closing we note that an effective equation of state $w_Q \simeq -1$ for
the present time, as the one we find, is favored by the available 
data \cite{efs}. 
Unfortunately this makes it even harder to observationally 
distinguish the scalar field contribution from the pure cosmological 
constant case.

\acknowledgements
We would like to thank Bobby Acharya, David Bailin, Tiago Barreiro, 
Philippe Brax, Beatriz de Carlos, Kiwoon Choi, Sacha Davidson, David Lyth, Cindy Ng, 
Massimo Pietroni and Fernando Quevedo for useful comments and discussions.
EJC is supported by PPARC, NJN by the Funda\c{c}\~{a}o para a Ci\^{e}ncia e a
Tecnologia (Portugal) and FR by the Ministero
dell'Universit\`a e della Ricerca Scientifica (Italy).

\end{document}